\documentclass[%
aps,
prl,
amsmath,amsfonts,amssymb,
reprint,
floatfix,
nofootinbib,
]{revtex4-2}

\usepackage{graphicx}
\usepackage{dcolumn}
\usepackage{bm}
\usepackage{mathptmx}

\usepackage{xcolor}
\definecolor{teal}{rgb}{0,0.5,0.5}
\usepackage{soul}

\usepackage[caption=false]{subfig}
\usepackage{anyfontsize}

\allowdisplaybreaks

\begin{document}

\title{Exact Screening-Ranged Expansions for Many-Body Electrostatics}

\author{Sergii V. Siryk}
\email{accandar@gmail.com}
\affiliation{CONCEPT Lab, Istituto Italiano di Tecnologia, Via E. Melen 83, 16152, Genova, Italy}
\author{Walter Rocchia}
\email[W. Rocchia \emph{(corresponding co-author)}: ]{walter.rocchia@iit.it}
\affiliation{CONCEPT Lab, Istituto Italiano di Tecnologia, Via E. Melen 83, 16152, Genova, Italy}


\begin{abstract}
\noindent
We present an exact many-body framework for electrostatic interactions among $N$ arbitrarily charged spheres in an electrolyte, modeled by the linearized Poisson--Boltzmann equation. Building on a spectral analysis of nonstandard Neumann--Poincar\'e-type operators introduced in a companion mathematical work~\textcolor{red}{\cite{supplem_pre_math}}, we construct convergent screening-ranged series for the potential, interaction energy, and forces, where each term is associated with a well-defined Debye--H\"uckel screening order and can be obtained evaluating an analytical expression rather than numerically solving an infinitely dimensional linear system. This formulation unifies and extends classical and recent approaches, providing a rigorous basis for electrostatic interactions among heterogeneously charged particles (including Janus colloids) and yielding many-body generalizations of analytical explicit-form results previously available only for two-body systems. The framework captures and clarifies complex effects such as asymmetric dielectric screening, opposite-charge repulsion, and like-charge attraction, which remain largely analytically elusive in existing treatments. Beyond its fundamental significance, the method leads to numerically efficient schemes, offering a versatile tool for modeling colloids and soft/biological matter in electrolytic solution.
\end{abstract}

\maketitle

\emph{\textbf{Introduction.}} 
Electrostatic interactions control processes ranging from biomolecular recognition and binding to colloidal stability and self-assembly~\cite{SheinermanNorelHonig,Blossey2023,our_jcp,our_jpcb,BarrLui_2014,JPCBSIYR}. In soft and biological matter, they are often described within continuum electrostatics, where solutes and solvent are modeled as linear dielectrics~\cite{BesleyACR}. A central workhorse is the mean-field Poisson--Boltzmann equation (PBE)~\cite{Cisneros_ChemRev,Blossey2023}; its linearization (LPBE) underlies Debye--H\"uckel (DH) theory and is widely used to describe weakly charged systems and small electrostatic potentials~\cite{our_jpcb, our_jcp, Derb2, Derb4, BMP22}. Despite its formal linearity, LPBE-based descriptions remain informative even for highly charged solutes at separations larger than the Debye length, especially when charge renormalization or effective screening parameters are introduced~\cite{TrizPRL, Triz1, ST1, Boon2015, our_jcp, our_jpcb, Krishnan2017, SchlaichHolm, BritoDenton, BoonJanus2010, Alex1, Fish, JanNetz09, Yu2021}. As a result, LPBE continues to play a central role in coarse-grained modeling and motivates ongoing analytical and computational work~\cite{Ponce2024,our_jcp,our_jpcb,Yu3,Yu2021,BritoDenton,Tabrizi1,FilStar_jpcb,BesleyACR,Derb2,Derb4,DerbFil2017,LinQS,Jha1,AddisonSmithCooper2023,Silva1,SCW,BoWangjcp,RMDP,FinBar2016,MAR1,KMRR,BSBBC,FBYJBH,Ether2018,FolOnu,BMP22,JNQS,MajOsh2025,CMWA,DGKF_2025,DiFlorio2025NextGenPB,FEM-BEM_diffuse_interface,WK2}.  

For a single dielectric sphere containing several point charges and located in an electrolyte (described in~\cite{Kirkwood1934}) the LPBE admits an explicit solution, but already for two interacting spheres ($N=2$) with different dielectric constants mutual polarization makes the problem highly nontrivial~\cite{Fish, Derb2, BesleyACR, LCSB, our_jcp, McClurg, Yu3}. Already in the Poisson limit of zero ionic strength, coupled multipolar expansions subject to infinitely many boundary constraints on the coefficients severely hinder analytical progress~\cite{GWJXL,our_jpcb,LCSB,BSL2014,Yu2019,DGC_softmat}. As emphasized in Ref.~\cite{BSL2014}, ``An implicit series expansion is known for the system of two dielectric spheres. For more than two dielectric spheres, numerical treatment is required." Since then, image-based multiple-scattering formalisms~\cite{Qin2019,GXFQ,QPF,Freed2014}, Wigner-matrix-free approaches~\cite{Yu2019,Yu3,Yu2021}, and hybrid analytical-numerical schemes~\cite{GJLX,GWJXL,DGC_softmat,LSB2018, LinQS, HasStammI, Hassan_Stamm_jctc, Lindgren_jcis, FBYJBH} have substantially extended what can be done for two spheres, but a general explicit many-body ($N>2$) analytical treatment at finite ionic strength is still lacking. Ionic screening further increases the technical complexity~\cite{Fish,LLF,LinQS,DuanGan2025,Yu3}, limits the applicability of image-charge constructions~\cite{DuanGan2025}, and leaves key questions on like-charge attraction (LCA), opposite-charge repulsion (OCR), and asymmetric dielectric screening (ADS) largely unresolved~\cite{Yu2021,Derb2,LCSB,DGC_softmat,DuanGan2025}.

In this Letter, we address this gap within the LPBE/DH framework by developing an exact and explicit many-body formalism for $N$ arbitrarily charged dielectric spheres in an electrolyte. Using the spectral properties of nonstandard Neumann--Poincar\'e-type operators introduced in the companion mathematical work~\textcolor{red}{\cite{supplem_pre_math}}, we construct a convergent Neumann-type expansion for the multipole coefficients of the DH potentials and thereby obtain screening-ranged series for the potentials, interaction energies, and forces. Each term in these series is associated with a definite ``screening order" and contains products of DH factors $e^{-\kappa R_{i j}}/R_{i j}$ along interaction paths, providing a systematic and controllable many-body extension to the electrostatic part of the familiar Derjaguin–Landau–Verwey–Overbeek (DLVO) and to the unscreened Coulomb descriptions. The formalism applies to general fixed charge distributions, including heterogeneously charged and Janus-like particles, and yields explicit expressions that are straightforward to evaluate numerically. 

We illustrate the scope of our formalism by (i) deriving exact higher-order screened corrections for charged spheres, clarifying the conditions for LCA, OCR, and ADS, and (ii) constructing compact DLVO-like interaction expressions for Janus particles with controlled many-body and higher-multipole corrections. We highlight its value through two representative applications. 
For monopolar charged spheres, we recover the standard DLVO interaction as the leading screening order and obtain exact higher-order many-body corrections that generalize two-sphere Fisher--Levin--Li, McClurg--Zukoski, and related energy and force expressions~\cite{Fish,LLF,McClurg,FinBar2016,our_jcp,Yu3,Derb2,Derb4,BesleyACR,FilStar_jpcb}. 
Even in the long-studied two-sphere problem, this framework yields new analytical insight: it provides explicit conditions for the (non)occurrence of LCA and OCR, clarifies the role of dielectric contrast and ionic strength in ADS, and recovers recent image-charge-based polarization-energy results as a particular instance of the double-screened contribution~\cite{DGC_softmat,DuanGan2025,RPSA_2024,Chan}. 
For Janus-type particles with bipatchy surface charge, we derive compact DLVO-like expressions for effective monopole-monopole, monopole-dipole, and dipole-dipole interactions~\cite{GraafBoon2012,GLCBB_2025}, and discuss how higher screening orders systematically generate many-body and higher-multipole corrections. 
Taken together, these examples demonstrate that the proposed screening-ranged expansions provide a versatile analytical framework for electrostatic interactions in model colloids and coarse-grained biomolecular systems, directly amenable to efficient numerical implementation. 
Beyond the specific cases discussed here, the framework offers a controlled many-body expansion for screened electrostatics, organized in a compact, rapidly convergent series that can underpin effective interaction models in charged soft-matter systems, from patchy colloids to biomolecular assemblies; further examples, including particles with clouds of internal point charges, higher-order fixed multipoles, or inhomogeneous free surface distributions, are developed in the joint papers~\textcolor{red}{\cite{supplem_pre,supplem_pre_force}}.

\emph{\textbf{Problem statement.}} 
We consider a general system of $N$ non-overlapping spherical dielectric particles ($\Omega_i\subset\mathbb R^3$, $i\in\{1,\ldots,N\}$) immersed in an electrolytic solvent (e.g.~water with mobile ions) with dielectric constant $\varepsilon_\text{sol}$ and Debye screening length $\kappa^{-1}$. Each particle $\Omega_i$ is centered at $\mathbf x_i\in\mathbb R^3$ and is characterized by its dielectric constant $\varepsilon_i$ and radius $a_i$; $\rho_i^\text{f}(\mathbf r)$ denotes the free (fixed) charge density inside $\Omega_i$.  For any $i$, the electrostatic potential $\Phi_{\text{in},i}(\mathbf r)$ inside $\Omega_i$ (i.e.~for $r_i < a_i$, where $r_i = \|\mathbf r - \mathbf x_i\|$) satisfies the Poisson equation
\[
\Delta\Phi_{\text{in},i}(\mathbf r)= - \rho_i^\text{f}(\mathbf r)/(\varepsilon_i \varepsilon_0),
\]
while the corresponding potential $\Phi_{\text{out},i}$ in the solvent domain $\Omega_\text{sol}$ obeys the LPBE~\cite{our_jcp,our_jpcb}
\[
\Delta\Phi_{\text{out},i}(\mathbf r) - \kappa^2 \Phi_{\text{out},i}(\mathbf r) = 0 .
\]
By the superposition principle inherent in the DH description~\cite{Fish,Derb2}, the total electrostatic potential~is
\[
\Phi(\mathbf r) = 
\begin{cases}
\Phi_{\text{in},i}(\mathbf r), & \mathbf r \in \Omega_i, \\
\Phi_{\text{out}}(\mathbf r) := \sum_{i=1}^N \Phi_{\text{out},i}(\mathbf r), & \mathbf r \in \Omega_\text{sol}.
\end{cases}
\]

At each sphere-solvent interface, $\Phi$ satisfies transmission-type boundary conditions (BCs):
\begin{gather*}
\varepsilon_i (\mathbf n_i \cdot \nabla \Phi_{\text{in},i})\big|_{r_i\to a_i^-} - \varepsilon_\text{sol} (\mathbf n_i \cdot \nabla \Phi_{\text{out}})\big|_{r_i\to a_i^+} = \sigma_i^\text{f}/\varepsilon_0,\\
\Phi_{\text{in},i}\big|_{r_i \to a_i^-} =\Phi_{\text{out}}\big|_{r_i \to a_i^+} ,
\end{gather*}
where $\mathbf n_i$ is the outer unit normal and $\sigma_i^\text{f}$ is any free charge density on the boundary $\partial\Omega_i$ ($r_i = a_i$), with $r_i\to a_i^{\pm}$ denoting approach from the interior ($-$) or exterior ($+$) side. While other types of BCs are also possible~\textcolor{red}{\cite{supplem_pre}}, here we focus on transmission BCs \cite{CurLui_2021}, which represent a situation of primary interest in biomolecular electrostatics. We also impose the usual condition that the potential vanishes at infinity.

\emph{\textbf{Screening-ranged expansions for potentials.}}  
Within each particle $\Omega_i$, the potential is decomposed into a known Coulomb contribution generated by the fixed charge density and an induced harmonic correction, while in the solvent domain the homogeneous LPBE is solved. Using complex spherical harmonics $Y_n^m$ in local spherical coordinates centered at each sphere, we represent both the interior correction and the exterior DH potential as multipole expansions with unknown coefficients. Mutual polarization between particles is encoded by exact re-expansions for screened solid harmonics (of the type \eqref{Yu3_reexp}), which express the field generated by sphere $j$ in the local frame of sphere $i$.  
Imposing the transmission BCs at all interfaces then leads to a coupled global linear system for the unknown exterior multipole coefficients. In block form, this system can be written~as
\begin{equation}
(\mathbb I + \mathbb K)\Vec{\Tilde{\mathbb G}} = \Vec{\Tilde{\mathbb S}},
\label{global_lin_sys1}
\end{equation}
where $\Vec{\Tilde{\mathbb G}}$ collects all exterior multipole coefficients, $\Vec{\Tilde{\mathbb S}}$ gathers the source terms due to fixed charges ($\rho_i^\text{f}$, $\sigma_i^\text{f}$), $\mathbb I$ is the identity and $\mathbb K$ is a compact operator that couples different particles and multipoles through screened propagators proportional to $e^{-\kappa R_{i j}}/R_{i j}$ ($R_{i j}$ is the magnitude of vector $\mathbf R_{i j}=\mathbf x_j-\mathbf x_i$ connecting sphere centers). A detailed construction of $\mathbb K$ and $\Vec{\Tilde{\mathbb S}}$ is given in \textcolor{red}{\cite{supplem_pre_math,supplem_pre}} (see End Matter). The spectral analysis in \textcolor{red}{\cite{supplem_pre_math}} shows that the spectral radius of $\mathbb K$ considered in the relevant Hilbert space of multipole coefficients is strictly less than $1$. This guarantees that the inverse of $(\mathbb I + \mathbb K)$ admits a convergent Neumann-type operator series representation,
\begin{equation}
\Vec{\Tilde{\mathbb G}} = \sum\nolimits_{\ell=0}^{+\infty}(-1)^\ell \mathbb K^\ell \Vec{\Tilde{\mathbb S}} ,
\label{G_neumann_series0}
\end{equation}
and allows us to decompose the solution for the multipole coefficients into screening orders. Each power $\mathbb K^\ell$ corresponds to sequences of $\ell$ screened couplings between particles and generates contributions proportional to products of DH screening factors $e^{-\kappa R_{i j}}/R_{i j}$ along paths of length $\ell$ in the interaction network. Projecting~\eqref{G_neumann_series0} componentwise leads to the expansions
\begin{equation}
\Tilde{\mathbf G}_i = \sum\nolimits_{\ell=0}^{+\infty}\Tilde{\mathbf G}_i^{(\ell)}, 
\qquad
\mathbf L_i = \sum\nolimits_{\ell=0}^{+\infty}\mathbf L_i^{(\ell)},
\label{GL_componentwise0}
\end{equation}
where $\Tilde{\mathbf G}_i^{(\ell)}$ and $\mathbf L_i^{(\ell)}$ are the $\ell$-th order screened contributions to the exterior and interior multipole coefficients of particle $i$, respectively (explicit formulas are collected in End Matter and in \textcolor{red}{\cite{supplem_pre,supplem_pre_math}}). In particular, $\ell=0$ terms describe local solvation response (Born/Kirkwood) that is independent of interparticle distances $R_{i j}$, $\ell=1$ corresponds to pairwise (DLVO-like) interactions, while $\ell\ge2$ terms incorporate many-body screened interactions in a systematic and absolutely convergent series ordered by the screening range. This ordering in terms of path length and screening strength provides a natural small parameter and a controllable truncation scheme for practical implementations.

\emph{\textbf{Screening-ranged expansions for energy and force.}} The above screening-ranged relations \eqref{GL_componentwise0} in turn give rise to the explicit construction of the corresponding screening-ranged expansions for the total electrostatic energy $\mathcal E$ and force $\mathbf{F}_i$ acting on an arbitrary particle~$i$:
\begin{align}
\mathcal E & = \sum\nolimits_{\ell=0}^{+\infty}\mathcal E^{(\ell)}, \label{energy_expansion_components_abs} \\
\mathbf{F}_i & = \sum\nolimits_{\ell=1}^{+\infty}\mathbf{F}_i^{(\ell)} \label{Force_expansion_vec}
\end{align}
(note that $\mathbf{F}_i^{(\ell)}$ is always zero at $\ell=0$, see \textcolor{red}{\cite{supplem_pre_force}}, so the sum in \eqref{Force_expansion_vec} starts at $\ell=1$). A general explicit construction of the addends $\mathcal E^{(\ell)}$ and $\mathbf{F}_i^{(\ell)}$ is provided in End Matter (see \eqref{energy_expansion_components_abs_details},~\eqref{Force_expansion_vec_xyz_components_ell}) and in \textcolor{red}{\cite{supplem_pre,supplem_pre_force}} for further technical details.

For instance, if the particles' free distributions are modeled by multipolar moments $\{\Hat L_{n m,i}\}$ co-centered with the particles, then, with the exclusion of self-energy terms (to prevent the divergence of $\mathcal E$, see e.g.~\cite{RAH1,our_jcp}), the non-screened ($\mathbf R_{i j}$-free) addend $\mathcal E^{(0)}$ corresponds to the sum of one-body solvation (``Born/Kirkwood") energies.  
Meanwhile,
\begin{equation}
\label{energy_expansion_components_interaction}
\mathcal E^\text{Int} \mathrel{:=} \mathcal E - \mathcal E^{(0)} = \sum\nolimits_{\ell=1}^{+\infty}\mathcal E^{(\ell)}
\end{equation}
represents the full interparticle interaction energy.  

In the particular case of monopolar charged spheres (e.g.~either fixed point charges in their centers or uniform surface charge densities), the first interaction addends, namely $\mathcal E^{(1)}$ of \eqref{energy_expansion_components_interaction} and $\mathbf{F}_i^{(1)}$ of \eqref{Force_expansion_vec}, reproduce the familiar pairwise electrostatic DLVO expressions. Higher-order ($\ell\ge2$) screened addends have previously been explicitly derived only for two-sphere ($N=2$) systems and in some approximate/special settings (see below and Refs.~\textcolor{red}{\cite{supplem_pre,supplem_pre_force}} for details).

The accurate quantification of $(\ell\ge2)$-screened terms is important because they reveal many-body and mutual polarization effects, significantly impacting the overall interaction landscape beyond leading pairwise DLVO and Coulombic contributions (see \cite{Fish,LLF,FinBar2016,our_jcp}).  
Thus, the many-body analytical formalism proposed here provides, to our knowledge, the first exact framework that quantifies all screening-ranged contributions ($\forall\ell\ge0$) and derives expressions for their terms in explicit form, which can be evaluated with arbitrary precision and without explicitly solving the system of equations for potential coefficients or surface polarization charges beforehand. Energies and forces are obtained on the same footing, which is essential for consistent dynamical and simulation studies. Remarkably, this formalism applies both to monopolar spheres and to arbitrary fixed charge distributions $\rho_i^\text{f}$ and~$\sigma_i^\text{f}$.

Concerning electrostatic forces, explicit and exact analytical expressions for force components in the LPBE framework were previously known only for specific two-sphere systems \textcolor{red}{\cite{supplem_pre_force}}. Here, based on the Maxwell stress tensor components derived in \textcolor{red}{\cite{supplem_pre_force}}, we obtain general analytical expressions for the total interaction force $\mathbf{F}_i$ on any particle $i$ (yielding, to our knowledge, the first analytical and explicit many-body force expressions for LPBE electrostatics of spheres beyond two-body geometries, see \textcolor{red}{\cite{supplem_pre_force}} for an overview). Combining these force relations with expansions \eqref{G_componentwise0} provides the explicit construction (see \eqref{Force_expansion_vec_xyz_components_ell}) of the desired screening-ranged force expansion \eqref{Force_expansion_vec} and exact quantification of all addends~$\mathbf{F}_i^{(\ell)}$.

The resulting screening-ranged many-body energy and force expansions \eqref{energy_expansion_components_abs}-\eqref{Force_expansion_vec} converge rapidly with $\ell$ \textcolor{red}{\cite{supplem_pre,supplem_pre_force}} and are easy to implement numerically.  
Thus, the first few addends are typically sufficient to accurately approximate the total energy $\mathcal E$ and force $\mathbf{F}_i$, and avoid the need for conventional approaches that require solving laborious systems of equations for potential coefficients or surface polarization charges. As documented in the companion works~\textcolor{red}{\cite{supplem_pre,supplem_pre_force}}, this rapid convergence holds for representative colloidal and biomolecular parameter sets. Since the most computationally demanding step in harmonics-expansion-based approaches is solving the linear system / inverting the corresponding matrix \cite{DY_2006}, the identification of easy-to-calculate first addends of \eqref{energy_expansion_components_abs}-\eqref{Force_expansion_vec} already constitutes a major computational simplification. In the companion papers \textcolor{red}{\cite{supplem_pre,supplem_pre_force}}, we present several examples and show that including up to the double-screened addends ($\ell=2$) already captures the key features of the full energy and force profiles, including LCA and OCR.

As meaningful examples, we apply our formalism to generalized Janus particles and to many interacting dielectric monopolar spheres. Already in the two-sphere case, the method provides new insight into ADS, OCR and LCA phenomena, yielding exact analytical results that improve upon known approximate or limited-case results.

While the theory has been developed for dielectric spheres in an electrolytic solution, it can be shown that, by substituting the coefficients from \eqref{GL_componentwise0} into \eqref{energy_expansion_components_abs}-\eqref{Force_expansion_vec}, one obtains well-posed addends and correctly reproduces the corresponding energy and force terms also in the $\kappa\to0$ limit. Similarly, the limit to infinity for the dielectric constants of some (or all) of the spheres will yield the behavior of conducting particles in the quasi-static regime.

\emph{\textbf{Application I: particles with centrally-located point charges.}} 
Consider $N$ spheres with point charges $q_i$ in their centers. For clarity, we detail here the energy and force addends of \eqref{energy_expansion_components_abs}-\eqref{Force_expansion_vec} up to $\ell=2$. Higher orders can be treated in the same systematic way within our framework~\textcolor{red}{\cite{supplem_pre}}. 
In this case, $\Hat L_{n m,i} = \kappa q_i \delta_{n,0} /(\sqrt{4\pi}\varepsilon_0\varepsilon_i)$ ($\delta_{\cdot,\cdot}$ is the Kronecker delta). Substituting these $\Hat L_{n m,i}$ into \eqref{L_componentwise0} and \eqref{energy_expansion_components_abs_center}, we recover the well-known expressions 
$
\mathcal E^{(0)} = \frac{1}{8\pi\varepsilon_0} \sum_{i=1}^N \frac{q_i^2}{a_i}\bigl(\frac{1}{(1+\kappa a_i)\varepsilon_\text{sol}}-\frac{1}{\varepsilon_i}\bigr),
$ 
corresponding to the sum of single-sphere solvation (``Born/Kirkwood") energies~\cite{our_jcp, Fish, CheDzu2008}, and 
$
\mathcal E^{(1)} = \frac{1}{8\pi\varepsilon_0\varepsilon_\text{sol}}\sum_{i=1}^N \sum_{j\ne i}^N\frac{q_i q_j e^{\Tilde a_i+\Tilde a_j-\Tilde R_{i j}}\kappa}{(1+\Tilde a_i)(1+\Tilde a_j)\Tilde R_{i j}},
$ 
which is the pairwise DLVO interaction energy. 

At $\ell=2$ our formalism yields the first many-body relation \eqref{energy_N_monopoles_2_screened_legendre}, which, to our knowledge, was previously available only in the special two-sphere ($N=2$) case. Indeed, the leading term of $\mathcal E^{(2)}\bigr|_{N=2}$ was derived by McClurg \& Zukoski~\cite{McClurg} (MZ approximation), while Fisher, Levin \& Li~\cite{Fish} obtained a simplified two-leading-terms form valid for small screened radii (FLL approximation, \eqref{FLL_approx_E2})\footnote{The key result of FLL theory, summarized in Eq.~\eqref{FLL_approx_E2}, is of particular importance in colloidal science, as for instance the modern vision of the non-additivity effects in binary mixtures is based on it~\cite{FinBar2016}.}. Only recently the exact two-sphere result was obtained, in~\cite{our_jcp}. None of these approaches, however, provides an exact analytical expression in the~$N>2$ case. 

For $\ell>2$, even less was known: e.g.~the leading-term approximation to $\mathcal E^{(3)}\bigr|_{N=2}$ was recently obtained in~\cite{Yu3} via an energy reciprocity argument, and its full explicit form later in~\cite{our_jcp}. Again, many-body generalizations remained missing. The present framework closes this gap: higher-order screened addends $\mathcal E^{(\ell)}$ follow directly from \eqref{L_componentwise0} and \eqref{energy_expansion_components_abs_center} using the above $\Hat L_{n m,i}$. Fully expanded forms are given in the companion paper~\textcolor{red}{\cite{supplem_pre}}. 

Importantly, this framework allows identifying the specific conditions for the (non)occurrence of less obvious phenomena, such as OCR and LCA. Indeed,
in the particular case of two spheres less polarizable than the surrounding medium ($\varepsilon_i\le\varepsilon_\text{sol}$) and arbitrary $q_i$, it turns out that odd-$\ell$ addends $\mathcal E^{(\ell)}\bigr|_{N=2}(R)$, such as DLVO ($\ell=1$), have the usual Coulombic behavior w.r.t.~$R$, showing opposite-charge attraction and like-charge repulsion, while even-$\ell$ addends with $\ell\ge2$ are repulsive irrespective of the charge. This structure permits OCR (through the competition between attractive odd-$\ell$ and repulsive even-$\ell$ contributions)\footnote{Indeed, when $q_1q_2<0$, competition between attractive odd-$\ell$ and repulsive even-$\ell$ addends can produce short-range OCR \textcolor{red}{\cite{supplem_pre}}. This corresponds to the ADS effect, an enhanced repulsion between dielectric bodies in media with high dielectric constant, observed in numerical modelling in~\cite{DY_2006,Yu2021}.} but forbids LCA. However, this is not the case when particles are more polarizable than the medium ($\varepsilon_i > \varepsilon_\text{sol}$) and $\kappa > 0$, where the ionic strength can play a decisive role on the occurrence of OCR and LCA. Despite the existing recent DH-level treatments on the topic (see e.g.~\cite{Yu2021} and references therein), to our knowledge, OCR in this regime has not been explicitly characterized before in the literature (see \textcolor{red}{\cite{supplem_pre}} for technical details).

\begin{figure*}[t!]
  \centering
  \includegraphics[width=0.95\textwidth]{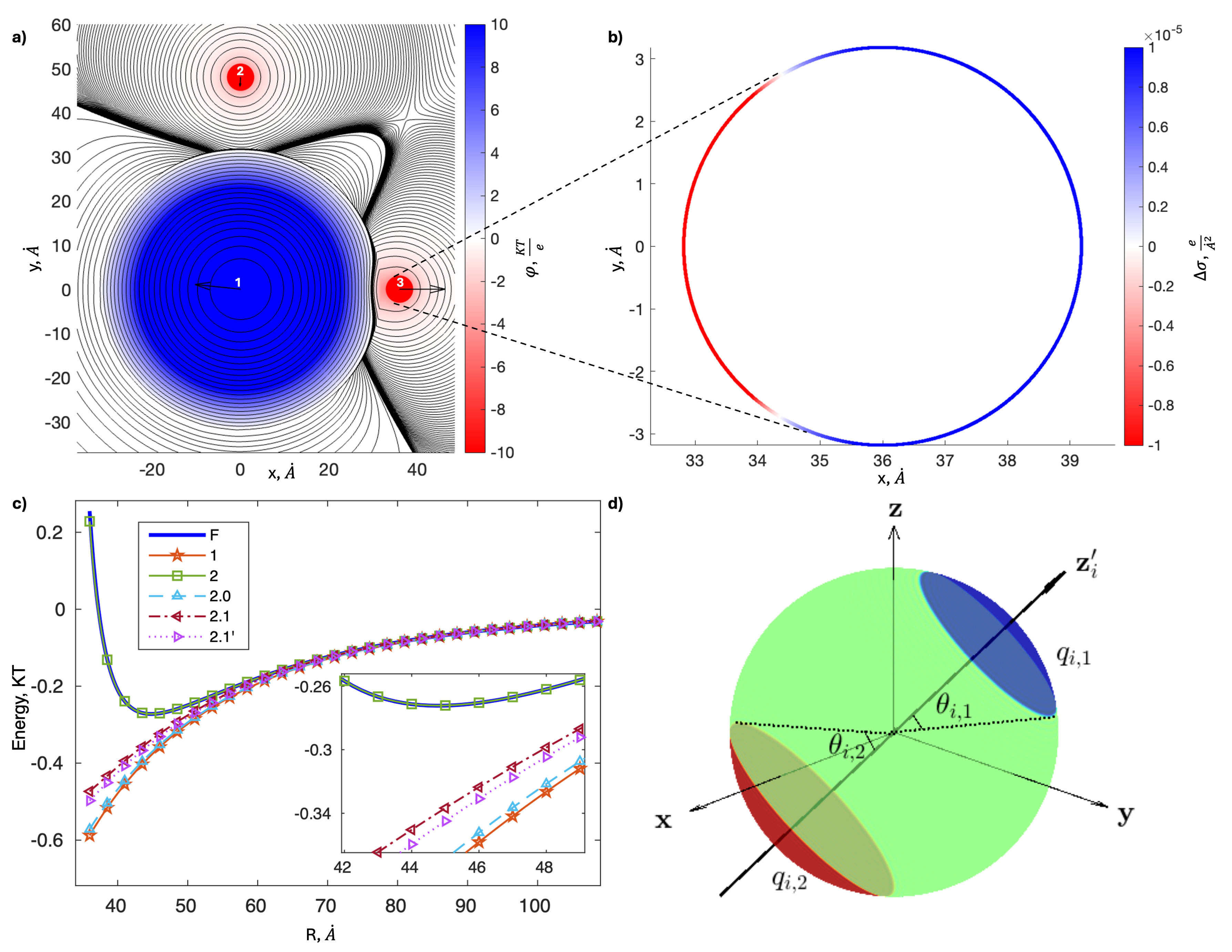}
  \caption{\textbf{a)} Electrostatic potential and iso-potential contour lines obtained for a three-sphere system. $\varepsilon_{1}=2$, $\varepsilon_{2, 3}=3$, $a_1+a_{2,3}=35$~\AA, $a_1=10 a_{2,3}$, $\varepsilon_\text{sol}=80$, $q_1=3e$, $q_{2,3} = -2e$, $\kappa^{-1}=8.07$~\AA. The arrows represent the forces acting on the spheres. Spheres $2$ and $3$ are identical, but located at a different distance from sphere 1.  The most significant correction introduced by the inclusion of higher order screened terms consists in an excess negative potential located inside sphere 3. This potential is more pronounced in the region facing sphere 1. This induces a difference in the particle polarization, reflected by the difference in surface charge distribution shown in panel \textbf{b)}.  \textbf{c)} Two spheres interaction energy w.r.t. $R=R_{1 2}$, $R\ge36$~\AA; $a_1+a_2=35$~\AA, $a_1=10 a_2$, $\varepsilon_1=2$, $\varepsilon_2=3$, $q_1=3 e$, $q_2 = -2 e$, $\varepsilon_\text{sol}=80$, $\kappa^{-1}=40$~\AA. Lines F, 1 and 2 depict the full $\mathcal E^\text{Int}$, $\mathcal E^{(1)}$ (i.e.~DLVO) and $\mathcal E^{(1)}+\mathcal E^{(2)}$, respectively. Lines 2.0, 2.1 and 2.1' correspond to one-term (i.e.~MZ), two-term and simplified two-term (i.e.~FLL) approximations of $\mathcal E^{(2)}$, respectively. (Embedded inset shows a close-up view). \textbf{d)} Example of $\sigma_i^\text{f}$ for a generalized Janus particle: free charges $q_{i,1}$ and $q_{i,2}$ are uniformly distributed over spherical caps of polar angles $[0,\, \theta_{i,1}]$ and $[\pi-\theta_{i,2},\,  \pi]$, respectively, while the intermediate surface (i.e.~$(\theta_{i,1},\, \pi-\theta_{i,2})$) has no fixed charge.}
  \label{4panels}
\end{figure*}

In panels \textbf{a}, \textbf{b} and \textbf{c} of Figure~\ref{4panels} the description of OCR on systems with two and three spheres with parameters typical of biophysical settings is presented. According to the DLVO approximation, both identical negative spheres in panel \textbf{a} are attracted by the central one. In contrast, the full formalism predicts that spheres 2 and 1 repel each other, due to the local effect of polarization charges which exceeds the Coulombic contribution. Similarly, in the two sphere case of panel \textbf{c}, one can see that while the DLVO energy $\mathcal E^{(1)}$, as well as the MZ and FLL approximations fail at short range, the inclusion of $\mathcal E^{(2)}$ is sufficient to capture the actual interaction profile.

In the $\kappa\to0$ limit, the relation \eqref{E2_point_charges_N_spheres_s_kappa_zero} is obtained. In the particular two-sphere case with $a_2\to0$, it reduces to the three-point image formula derived in the recent work~\cite{DuanGan2025}. Similarly, taking $\varepsilon_1\to+\infty$ and $a_i\to0$ for $i>1$ (i.e.~a cloud of point charges scattered around a charged conducting sphere), we reproduce the recent results of~\cite{RPSA_2024}. Unlike the formalism presented here, these image-charge-based approaches cannot be generalized to many dielectric spheres in electrolytic solution. Furthermore, asymptotic analysis of Eqs.~\eqref{energy_expansion_components_abs} and \eqref{L_componentwise0} in the $\kappa\to0$ two-sphere case recovers the same behavior already described (namely all odd-$\ell$ energy addends behave like the Coulombic term, while all even-$\ell$ addends behave like $\mathcal E^{(2)}$), when both spheres are either strictly less or more polarizable than the surrounding medium. This proves analytically that OCR cannot occur when $\varepsilon_i > \varepsilon_\text{sol}$ and LCA cannot occur when $\varepsilon_i < \varepsilon_\text{sol}$, rigorously answering the questions posed in the very recent (2025) work~\cite{DGC_softmat}. These results are consistent with recent effective-dipole-based computational studies~\cite{DGC_softmat} and empirical polarizability-volume arguments~\cite{Chan}.

As per the force expansion \eqref{Force_expansion_vec}, the $\ell=1$ contribution \eqref{Force_expansion_vec_xyz_components_ell} reproduces the DLVO force, while already at $\ell=2$ we obtain genuinely many-body terms \eqref{forces_double_screening_all_} with nontrivial angular couplings. These results were unknown except for the two-sphere case, where \eqref{monopoles_double_standard} yields \eqref{force_two_spheres_2scr_nonosmotic}, reducing further to \eqref{force_two_spheres_2scr_nonosmotic_Derb} under very restrictive assumptions: this last form overlaps with the approximate ion-molecule interaction result obtained in~\cite{Derb2}. 

\emph{\textbf{Application II: Janus particles.}} Our second example addresses the interaction of $N$ (generalized) bipatchy Janus particles. Here the fixed charge is located on the surface (see panel \textbf{d} of Fig.~\ref{4panels}; each particle has individual $\sigma_i^\text{f}$). Employing the Janus-specific densities $\sigma_i^\text{f}$ (see End Matter for details) in \eqref{energy_expansion_components_abs_surface} and \eqref{Force_expansion_vec_xyz_components_ell} directly generates the screening-ranged addends of the energy \eqref{energy_expansion_components_abs} and force \eqref{Force_expansion_vec} expansions. For concreteness we show in \eqref{Janus_expr} the explicit form of the starting energy addends at $\ell=\overline{0,1}$ in terms of the surface net free charge $q_i \mathrel{:=} q_{i,1}+q_{i,2}$ and the vector 
$$
\mathbf p_i \mathrel{:=} \tfrac{1}{2}\bigl(q_{i,1}(1+\cos\theta_{i,1})-q_{i,2}(1+\cos\theta_{i,2})\bigr) a_i \mathbf z_i^\prime 
$$ 
(noting that one would get $\mathbf p_i = (2 a_i)q_{i,1}\mathbf z_i^\prime$ if $q_{i,2}=-q_{i,1}$ and $\theta_{i,1},\theta_{i,2}\to0$, so that it is natural to interpret $\mathbf p_i$ as an effective physical dipole moment of the $i$-th Janus particle~\textcolor{red}{\cite{supplem_pre}}). By analogy with the pairwise interactions of monopolar spheres considered above, the terms made explicit in $\mathcal E^{(1)}$ of \eqref{Janus_expr} may be identified as effective monopole-monopole, monopole-dipole, and dipole-dipole interactions, respectively.~Exact analytical quantification of higher-order terms (which correspond to effective quadrupolar, octupolar, and higher multipolar interactions) can likewise be systematically obtained through the same expansion procedure. In practice, this hierarchy allows one to keep only the leading monopolar and dipolar terms when appropriate, while systematically incorporating higher multipoles when anisotropy plays a dominant~role. 

In recent years, there has been an increased interest in constructing anisotropic DLVO-like interaction descriptions for inhomogeneously/patchy charged particles, reflecting their growing importance in modern colloid and protein science -- see e.g.~the very recent (2025) work~\cite{GLCBB_2025} and references therein. A well-known extension of DLVO theory to describe the electrostatic pair interaction of dielectric particles with a substantial dipolar contribution to the surface charge, such as Janus particles, was proposed in 2012, \cite{GraafBoon2012}, where equal-sized Janus particles consisting of (generally unequally charged) hemispheres were considered. In this particular situation, one sees that our relation $\mathcal E^{(1)}$ of \eqref{Janus_expr} recovers and slightly extends the corresponding result of \cite{GraafBoon2012} (the only difference being that \cite[Eq.~(19b)]{GraafBoon2012} approximates our exact factor $\frac{e^{-\kappa R_{i j}}}{R_{i j}}\bigl(\kappa+\frac{1}{R_{i j}}\bigr)$ by $\frac{e^{-\kappa R_{i j}}}{R_{i j}^2}$ in monopole-dipole interactions).  

However, and importantly, unlike the theory described in \cite{GraafBoon2012}, our framework naturally goes beyond pairwise DLVO-like approximations: by setting $\ell>1$ in \eqref{energy_expansion_components_abs_surface}, we obtain exact explicit quantifications of higher-order screened and many-body contributions (detailed in~\textcolor{red}{\cite{supplem_pre}}). These contributions demonstrate that the proposed screening-ranged expansion formalism provides a rigorous and systematic route to explicitly and analytically capture anisotropic and collective electrostatic effects in patchy charged particles within the LPBE description.

In summary, we have introduced an exact many-body framework for electrostatics of dielectric spheres in electrolytic solution within the LPBE/DH description. By exploiting the spectral properties of Neumann--Poincar\'e-type operators introduced in \textcolor{red}{\cite{supplem_pre_math}}, we obtain convergent screening-ranged expansions for potentials, energies, and forces, organized by products of DH screening factors and directly applicable to arbitrary fixed charge distributions. The formalism recovers familiar DLVO interactions at leading order and provides explicit higher-order corrections that quantify many-body and mutual-polarization effects, clarify the conditions for ADS and LCA/OCR, and connect to recent image-charge and effective-dipole results~\cite{Fish,LLF,FinBar2016,our_jcp,DGC_softmat,DuanGan2025,RPSA_2024,Chan,Qin2019,GXFQ,QPF,Freed2014}. Our treatment of Janus-type particles further shows how anisotropic and patchy interactions can be captured analytically and extended beyond pairwise DLVO-like descriptions~\cite{GraafBoon2012,GLCBB_2025}. These results supply a compact, systematically improvable building block for modeling electrostatic interactions in colloidal and soft/biological systems, and they suggest natural extensions to more complex geometries and beyond-linear electrostatic theories. Because the expansions are ordered by screening range and interaction path length, they can be truncated with a clear physical error control, making them particularly suitable as analytical input for coarse-grained models, simulations, and the interpretation of nonadditive electrostatic effects. We anticipate that the resulting screening-ranged expansions for potentials, energies, and forces will provide a useful analytical backbone for reduced models and simulations, and for interpreting nonadditive interactions, OCR/LCA, and anisotropic colloids in experiment.

\emph{Data availability.} Codes/data can be found in~\textcolor{red}{\cite{our_github_rep}}.

\section{Acknowledgments}
\label{sec:Acknowledgements}
We acknowledge the financial support from the European 
Union - NextGenerationEU and the Ministry of University and Research (MUR), 
National Recovery and Resilience Plan (NRRP): 
Research program CN00000013 “National Centre for HPC, 
Big Data and Quantum Computing”, CUP: J33C22001180001, funded by the D.D. n.1031, 17.06.2022 and Mission 4, Component 2, Investment 1.4 - Avviso “Centri Nazionali” - D.D. n. 3138, 16 December 2021.

\appendix
\begin{center}
\textbf{END MATTER}

\textbf{Appendix: Technical details and notation}
\end{center}
We use the standard modified spherical Bessel functions $k_n(x) \mathrel{:=} \sqrt{2/\pi}K_{n+1/2}(x)/\sqrt{x}$,  $i_n(x) \mathrel{:=} \sqrt{\pi/2}I_{n+1/2}(x)/\sqrt{x}$.

\paragraph{\textbf{Derivation of screening-ranged expansions for potentials.}} $\Phi_{\text{in},i}$ admits~\cite{our_jcp,our_jpcb} decomposition $\Phi_{\text{in},i} = \Hat\varPhi_{\text{in},i} + \Tilde\Phi_{\text{in},i}$, where
$\Hat\varPhi_{\text{in},i}(\mathbf r)$ is the Coulombic potential for $\rho_i^\text{f}$, while $\Delta\Tilde\Phi_{\text{in},i}=0$. We express the unknown $\Tilde\Phi_{\text{in},i}$ and $\Phi_{\text{out},i}$ (where $|\Tilde\Phi_{\text{in},i}|<\infty$ as $r_i\to0^+$, $\Phi_{\text{out},i}\to0$ as $r_i\to+\infty$) via spherical harmonics~$Y^m_n$:
\begin{subequations}
\label{Lin_eqs_Phi_in_out}
\begin{gather}
\Tilde\Phi_{\text{in},i}(\mathbf r) = \sum\nolimits_{n,m}L_{n m,i} \Tilde r_i^n Y_n^m(\Hat{\mathbf r}_i), \label{Lin_eqs_Phi_in} \\
\Phi_{\text{out},i}(\mathbf r) = \sum\nolimits_{n,m} \Tilde G_{n m,i} \Upsilon_{n,i} k_n(\Tilde r_i) Y_n^m(\Hat{\mathbf r}_i) , \label{Lin_eqs_Phi_out}
\end{gather}
\end{subequations}
where $\mathbf r_i = \mathbf r - \mathbf x_i$, $\Tilde r_i \mathrel{:=} \kappa r_i$, $\Tilde a_i \mathrel{:=} \kappa a_i$, $\Hat{\mathbf r}_i \mathrel{:=} \mathbf r_i/r_i$, $\sum_{n,m} \mathrel{:=}  \sum_{n=0}^{+\infty} \sum_{m=-n}^n$, scaling $\Upsilon_{n,i} \mathrel{:=} \left((2 n+1) k_n(\Tilde a_i) a_i\right)^{-1}$; angles of $\Hat{\mathbf r}_i$ are measured in the parallel local coordinate frames of particles~\textcolor{red}{\cite{supplem_pre}}. We expand $\Hat\varPhi_{\text{in},i}(\mathbf r)$ (as $r_i \to a_i^-$) and $\sigma_i^\text{f}$ via $\Hat\varPhi_{\text{in},i} (\mathbf r) = \sum\nolimits_{n,m} \Hat L_{n m,i} \Tilde r_i^{-n-1} Y_n^m(\Hat{\mathbf r}_i)$, $\sigma_i^\text{f}(\Hat{\mathbf r}_i) = \sum_{n,m} \sigma_{n m,i}^\text{f} Y_n^m(\Hat{\mathbf r}_i)$, where $\Hat L_{n m,i} = \frac{\kappa}{(2 n+1)\varepsilon_i \varepsilon_0}\int_{\Omega_i} \rho_i^\text{f}(\mathbf r_i) \Tilde r_i^n Y_n^m(\Hat{\mathbf r}_i)^\star d\mathbf r_i$, $\sigma_{n m,i}^\text{f} = a_i^{-2} \oint_{\partial\Omega_i} \sigma_i^\text{f}(\Hat{\mathbf s}_i) Y_n^m(\Hat{\mathbf s}_i)^\star d s_i$. Unknown coefficients $L_{n m,i}$ and $\Tilde G_{n m,i}$ of \eqref{Lin_eqs_Phi_in_out} are determined from BCs. To represent all quantities via the same set $\{Y_n^m(\Hat{\mathbf r}_i)\}$ (for ensuring self-consistent treatment of mutual polarization), we treat $k_n(\Tilde r_j) Y_n^m(\Hat{\mathbf r}_j)$ of \eqref{Lin_eqs_Phi_out} for $j\ne i$ by employing fully-analytical re-expansions advanced in the recent Wigner-matrix-free formalism~\cite{Yu3,Yu2021}:
\begin{gather}
k_L(\Tilde r_j) Y_L^M(\Hat{\mathbf r}_j) = \sum\nolimits_{l_1,m_1}\mathcal H_{l_1 m_1}^{L M}(\mathbf{R}_{i j}) i_{l_1}(\Tilde r_i) Y_{l_1}^{m_1}(\Hat{\mathbf r}_i),\label{Yu3_reexp}
\end{gather}
where $\mathcal H_{l_1 m_1}^{L M}$ handles re-expansion between local frames\footnote{Other treatments of offside ($j\ne i$) contributions exist (Wigner rotation matrices, numerical iterative re-expansion, Lebedev quadratures, etc.~-- see~\textcolor{red}{\cite{supplem_pre}}), but they hinder a rigorous analytical study and are irrelevant for building an exact fully-analytical solution.}~\textcolor{red}{\cite{supplem_pre_math}}. Applying \eqref{Lin_eqs_Phi_in_out}-\eqref{Yu3_reexp} to BCs and eliminating the internal coefficients $L_{n m,i}$ (see \textcolor{red}{\cite{supplem_pre,supplem_pre_math}} for details), we obtain \eqref{global_lin_sys1} w.r.t.~the global block ``vector" $\Vec{\Tilde{\mathbb G}} = (\Tilde{\mathbf G}_1, \ldots, \Tilde{\mathbf G}_N)^T$ consisting of (infinite-size) vectors $\Tilde{\mathbf G}_i \mathrel{:=} \{\Tilde G_{n m,i}\}_{0\le\left|m\right|\le n}$. Inverting $\mathbb I + \mathbb K$ via an operator Neumann-type series\footnote{This inversion is justified by spectral properties of $\mathbb K$ proved in \textcolor{red}{\cite{supplem_pre_math}} (note that traditional spectra-localization-based methods fail here due to row/column norms exceeding $1$, in general, and the fact that $\mathbb K$ consists of infinite-sized blocks).} gives \eqref{G_neumann_series0}. Thanks to the structure of $\mathbb K$, each $\Vec{\Tilde{\mathbb G}}^{(\ell)}$ corresponds to screening order $\ell$, proportional to products of DH screening factors $e^{-\kappa R_{i j}}/R_{i j}$, and this entails relations~$\forall i$:
\begin{align}
\!\!\!\!
& \qquad\qquad\qquad\qquad \Tilde{\mathbf G}_i = \sum\nolimits_{\ell=0}^{+\infty}\Tilde{\mathbf G}_i^{(\ell)}, \label{G_componentwise0} \\
& \Tilde{\mathbf G}_i^{(0)}\!=\Tilde{\mathbf S}_i , \ \Tilde{\mathbf G}_i^{(1)} \!=\! -\!\sum\nolimits_{j\ne i}^N\!\mathsf K_{i j} \Tilde{\mathbf S}_j , \  \Tilde{\mathbf G}_i^{(2)} \!=\! \sum\nolimits_{j=1}^N\!\sum\nolimits_{k\ne i, j}^N\!\mathsf K_{i k} \mathsf K_{k j} \Tilde{\mathbf S}_j , \notag\\
& \Tilde{\mathbf G}_i^{(\ell)} = (-1)^\ell \sum\nolimits_{j=1}^N \, \sum\nolimits^N_{k_{\ell-1}\ne i} \mathsf K_{i k_{\ell-1}} \sum\nolimits_{k_{\ell-2}\ne k_{\ell-1}}^N \mathsf K_{k_{\ell-1} k_{\ell-2}} \notag\\ 
&\qquad\quad\times\cdots \times \sum\nolimits_{k_1\ne k_2, j}^N \mathsf K_{k_2 k_1} \mathsf K_{k_1 j} \Tilde{\mathbf S}_j \, ,\notag
\end{align}
where matrix $\mathsf K_{i j}$ and vector $\Tilde{\mathbf S}_i$ are block components of $\mathbb K$ and $\Vec{\Tilde{\mathbb S}}$ \textcolor{red}{\cite{supplem_pre_math,supplem_pre}}; $\Tilde{\mathbf G}_i^{(0)}$ is independent of any $R_{i j}$, $\Tilde{\mathbf G}_i^{(1)}$ contains elements proportional to factors $\frac{e^{-\Tilde R_{i j}}}{R_{i j}}$ ($\Tilde R_{i j}\mathrel{:=}\kappa R_{i j}$, $j\ne i$), $\Tilde{\mathbf G}_i^{(2)}$ --- to $\frac{e^{-\Tilde R_{i k}}}{R_{i k}} \frac{e^{-\Tilde R_{k j}}}{R_{k j}}$ ($k\ne i,j$), and so forth~\textcolor{red}{\cite{supplem_pre}}. Similarly, for the vector of internal coefficients $\mathbf L_i = \{L_{n m,i}\}_{0\le\left|m\right|\le n}$ we build 
\begin{align}
& \qquad\qquad\quad \mathbf L_i = \sum\nolimits_{\ell=0}^{+\infty}\mathbf L_i^{(\ell)}, \label{L_componentwise0}\\
\mathbf L_i^{(0)} &= \mathsf C_i \Tilde{\mathbf G}_i^{(0)} + \mathbf E_i , \quad
\mathbf L_i^{(\ell)} = \sum\nolimits_{j\ne i}^N \mathsf P_{i j} \Tilde{\mathbf G}_j^{(\ell-1)}, \ \ell\ge1 , \notag
\end{align}
with $\Tilde{\mathbf G}_i^{(\ell)}$ from \eqref{G_componentwise0}; specific matrices $\mathsf P_{i j}$ ($\propto e^{-\kappa R_{i j}}/R_{i j}$), $\mathsf C_i$, and vector $\mathbf E_i$ are derived and recorded in detail in~\textcolor{red}{\cite{supplem_pre}}. The elements of $\Tilde{\mathbf G}_i^{(\ell)}$ and $\mathbf L_i^{(\ell)}$ are detailed in~\textcolor{red}{\cite{supplem_pre}}, with rigorous element-wise absolute convergence proved in~\textcolor{red}{\cite{supplem_pre_math}}.

\paragraph{\textbf{Spectral properties of $\mathbb K$.}} Coefficients $\Tilde G_{n m,i}$ of the DH potentials form infinite square-summable sequences belonging to the $l^2$ Hilbert space~\textcolor{red}{\cite{supplem_pre_math}}. We require that spheres are non-overlapping, but do not make any further additional assumption on $\kappa$, $a_i$, $R_{i j}$, or on their ratios. We then prove in \textcolor{red}{\cite{supplem_pre_math}} that operator $\mathbb K \colon \pmb l^2 \to \pmb l^2$, where $\pmb l^2 \mathrel{:=} \bigoplus_{i=1}^N l^2(\{ \Tilde G_{n m,i} \}_{0\le |m|\le n})$, is compact and discuss the spectral radius bound $r(\mathbb K)<1$ to prove that the operator series in~\eqref{G_neumann_series0} converges.

\paragraph{\textbf{General construction of the screened energy addends in~\eqref{energy_expansion_components_abs}.}} 
First assuming for simplicity $\forall\sigma_i^\text{f}=0$ we have $\mathcal E = \frac{1}{2}\int_{\mathbb R^3} \rho^\text{fixed}(\mathbf r) \Phi(\mathbf r) d \mathbf r = \frac{1}{2}\sum_{i=1}^N\int_{\Omega_i} \rho_i^\text{f}(\mathbf r) \Phi_{\text{in},i}(\mathbf r) d \mathbf r$. The predetermined non-screened ($\mathbf R_{i j}$-free) $\Hat\varPhi_{\text{in},i}$-contributions (if any) are to be ascribed to $\mathcal E^{(0)}$, while for $\Tilde\Phi_{\text{in},i}$-contributions we have $\frac{1}{2}\sum_{i=1}^N\int_{\Omega_i} \rho_i^\text{f}(\mathbf r) \Tilde\Phi_{\text{in},i}(\mathbf r) d\mathbf r = \frac{1}{2}\sum_{i=1}^N\frac{\varepsilon_i \varepsilon_0}{\kappa}\sum_{n,m}(2 n+1)\Hat L_{n m,i}^\star L_{n m,i}$; using $L_{n m,i} = \sum_{\ell=0}^{+\infty}L_{n m,i}^{(\ell)}$ yielded by \eqref{L_componentwise0} we~get
\begin{subequations}
\label{energy_expansion_components_abs_details}
\begin{align}
\mathcal E^{(\ell)} & = \tfrac{1}{2}\sum\nolimits_{i=1}^N\tfrac{\varepsilon_i \varepsilon_0}{\kappa}\sum\nolimits_{n,m}(2 n+1)\Hat L_{n m,i}^\star L_{n m,i}^{(\ell)} . \label{energy_expansion_components_abs_center}
\intertext{Likewise in the case of $\sigma_i^\text{f}\ne0$ but $\forall\rho_i^\text{f}=0$, from $\mathcal E = \frac{1}{2}\sum_{i=1}^N\oint_{\partial\Omega_i} \sigma_i^\text{f}(\Hat{\mathbf s}) \Phi(\Hat{\mathbf s}) d s$ we obtain addends}
\mathcal E^{(\ell)} & = \tfrac{1}{2}\sum\nolimits_{i=1}^N a_i^2 \sum\nolimits_{n,m} \sigma_{n m,i}^{\text{f} \ \star} \Tilde a_i^n L_{n m,i}^{(\ell)} . \label{energy_expansion_components_abs_surface}
\end{align}
\end{subequations}
Finally, the case of both nonzero $\rho_i^\text{f}$ and $\sigma_i^\text{f}$ can be handled using the ones discussed above.

\paragraph{\textbf{General construction of the screened force addends in~\eqref{Force_expansion_vec}.}} 
We have (for derivation details see~\textcolor{red}{\cite{supplem_pre_force}})
\begin{equation}
\label{Force_expansion_vec_xyz_components_ell}
(\mathbf{F}_i^{(\ell)})_x = \operatorname{Re} \aleph_i^{(\ell)} , \quad 
(\mathbf{F}_i^{(\ell)})_y = \operatorname{Im} \aleph_i^{(\ell)} , \quad
(\mathbf{F}_i^{(\ell)})_z = \beth_i^{(\ell)}
\end{equation}
where the corresponding $\ell$-screened quantities $\aleph_i^{(\ell)}$, $\beth_i^{(\ell)}$~are
\begin{align*}
&\aleph_i^{(\ell)} = \varepsilon_0\varepsilon_\text{sol}\!\sum\nolimits_{n,m}\!\sqrt{\!\tfrac{(n-m+1) (n-m+2)}{(2 n+1) (2 n+3)}} \sum\nolimits_{k=0}^\ell \bigl( B_{n+1,m-1;i}^{(k)} A_{n,m;i}^{(\ell-k)}{}^\star \\
&\quad - \Tilde a_i^2 \Psi_{n+1,m-1;i}^{(k)}\Psi_{n,m;i}^{(\ell-k)}{}^\star\bigr) \quad\text{and} \\ 
&\beth_i^{(\ell)} = \varepsilon_0\varepsilon_\text{sol}\sum\nolimits_{n,m} \sqrt{\!\tfrac{(n-m+1) (n+m+1)}{(2 n+1) (2 n+3)}} \sum\nolimits_{k=0}^\ell \bigl(B_{n+1,m;i}^{(k)} A_{n,m;i}^{(\ell-k)}{}^\star \\
&\quad - \Tilde a_i^2\Psi_{n+1,m;i}^{(k)}\Psi_{n,m;i}^{(\ell-k)}{}^\star\bigr), \quad\text{where $\forall k=\overline{0,\ldots,\ell}$ it is defined} \\
&A_{n,m;i}^{(k)} = \Xi_{n,m;i}^{(k)} - n\Psi_{n,m;i}^{(k)}, \qquad B_{n,m;i}^{(k)} = \Xi_{n,m;i}^{(k)} + (n+1)\Psi_{n,m;i}^{(k)} ,\\
&\Psi_{n,m;i}^{(k)} = k_n(\Tilde a_i) G_{n m,i}^{(k)} + i_n(\Tilde a_i)\!\sum\nolimits_{j=1, j\ne i}^N \! \sum\nolimits_{L,M}\mathcal H_{n m}^{L M}\!(\mathbf R_{i j}) G_{L M,j}^{(k-1)}\!, \\
&\Xi_{n,m;i}^{(k)} = \bigl(n k_n(\Tilde a_i) - \Tilde a_i k_{n+1}(\Tilde a_i)\bigr) G_{n m,i}^{(k)} + \bigl(n i_n(\Tilde a_i) + \Tilde a_i i_{n+1}(\Tilde a_i)\bigr)\\
&\quad\times \sum\nolimits_{j=1,\, j\ne i}^N\sum\nolimits_{L,M}\mathcal H_{n m}^{L M}(\mathbf R_{i j}) G_{L M,j}^{(k-1)}
\end{align*} 
(sums involving $G_{L M,j}^{(k-1)}$ in $\Psi_{n,m;i}^{(k)}$ and $\Xi_{n,m;i}^{(k)}$ are omitted at $k=0$, and $G_{n m,i}^{(k)} = \Tilde G_{n m,i}^{(k)} \Upsilon_{n,i}$ due to the scale $\Upsilon_{n,i}$ defined above). Total force $\mathbf{F}_i$ admits decomposition $\mathbf{F}_i = \Breve{\mathbf{F}}_i + \mathring{\mathbf{F}}_i$, where $\Breve{\mathbf{F}}_i$ and $\mathring{\mathbf{F}}_i$ represent the conventional electrostatic and the osmotic parts of Maxwell stress tensor \textcolor{red}{\cite{supplem_pre_force}}; in the expressions above the $\mathring{\mathbf{F}}_i$-components are determined by the $\Psi\Psi^\star$-type products.

\begin{widetext}
\paragraph{\textbf{Spheres with centrally-located point charges $q_i$.}} 
Here $\Hat L_{n m,i} = \kappa q_i \delta_{n,0} /(\sqrt{4\pi}\varepsilon_0\varepsilon_i)$, so we obtain elaborating on \eqref{energy_expansion_components_abs_center}:
\begin{subequations}
\label{energy_N_monopoles_2_screened_legendre}
\begin{align}
\!\!\!\!\!\!\! \mathcal E^{(2)} & = \frac{-\kappa}{8\pi\varepsilon_0\varepsilon_\text{sol}}\!\sum_{i=1}^N \!\frac{q_i e^{\Tilde a_i}}{1+\Tilde a_i}\!\sum_{j=1, j\ne i}^N \, \sum_{p=1, p\ne j}^N \!\frac{q_p e^{\Tilde a_p}}{1+\Tilde a_p} \sum_{l=0}^{+\infty} \frac{\! (\varepsilon_j-\varepsilon_\text{sol}) l \Tilde a_j^{-1} i_l(\Tilde a_j) -\varepsilon_\text{sol}i_{l+1}(\Tilde a_j) \!\! }{(\varepsilon_j-\varepsilon_\text{sol}) l \Tilde a_j^{-1} k_l(\Tilde a_j) + \varepsilon_\text{sol} k_{l+1}(\Tilde a_j)}\!(2 l+1) k_l(\Tilde R_{i j}) k_l(\Tilde R_{j p}) P_l(\cos\gamma_{j i,j p}) \label{energy_N_monopoles_2_screened_legendre_1} \\
\!\!\!\!\!\!\! & = \frac{\kappa}{16\pi\varepsilon_0\varepsilon_\text{sol}}\sum_{i=1}^N \frac{q_i e^{\Tilde a_i}}{1+\Tilde a_i} \sum_{j=1,\, j\ne i}^N \; \sum_{p=1,\, p\ne j}^N \frac{q_p e^{\Tilde a_p}}{1+\Tilde a_p} \frac{e^{-\Tilde R_{i j}}}{\Tilde R_{i j}} \frac{e^{-\Tilde R_{j p}}}{\Tilde R_{j p}} \biggl(\!d_{0,j} \!-\! 3 d_{2,j} \Bigl(\!1\!+\!\frac{1}{\Tilde R_{i j}}\!\Bigr)\!\Bigl(\!1\!+\!\frac{1}{\Tilde R_{j p}}\!\Bigr)\!\cos\gamma_{j i,j p}\!\biggr)\! + \bigl/l\ge2\;\ terms\bigr/ , \label{energy_N_monopoles_2_screened_legendre_2}
\end{align}
\end{subequations}
where \eqref{energy_N_monopoles_2_screened_legendre_2} merely shows an expanded form of the $l=0, 1$ addends of \eqref{energy_N_monopoles_2_screened_legendre_1} after explicit substitution of the Bessel functions into them, $\gamma_{j i,j p}$ is the angle between $\Hat{\mathbf R}_{j i}$ and $\Hat{\mathbf R}_{j p}$, and $d_{0,i} \mathrel{:=} e^{2\Tilde a_i} \frac{\Tilde a_i-1}{\Tilde a_i+1}+1$, $d_{2,i} \mathrel{:=} e^{2\Tilde a_i}\frac{(\varepsilon_i+2\varepsilon_{\mathsf{sol}})(\Tilde a_i-1)-\varepsilon_{\mathsf{sol}}\Tilde a_i^2}{(\varepsilon_i+2\varepsilon_{\mathsf{sol}})(1+\Tilde a_i)+\varepsilon_{\mathsf{sol}}\Tilde a_i^2}+1$. If $N=2$ and $\Tilde a_1$, $\Tilde a_2$ are small enough (small/weakly-screened particles) so that $d_{0,i} \approx \frac{2}{3}\Tilde a_i^3$, $d_{2,i} \approx \frac{2(\varepsilon_i-\varepsilon_\text{sol})}{3(\varepsilon_i+2\varepsilon_\text{sol})}\Tilde a_i^3$, then \eqref{energy_N_monopoles_2_screened_legendre_2} (omitting $l\ge2$-terms)~yields
\begin{gather}
\mathcal E^{(2)}\bigr|_{N=2} \approx \frac{\kappa e^{-2\Tilde R_{1 2}}}{8\pi\varepsilon_0\varepsilon_\text{sol} \Tilde R_{1 2}^2}  \biggl[\frac{q_1^2 e^{2\Tilde a_1} \Tilde a_2^3}{(1+\Tilde a_1)^2} \biggl( \frac{1}{3} + \Bigl(1+\frac{1}{\Tilde R_{1 2}}\Bigr)^2 
 \frac{\varepsilon_\text{sol}-\varepsilon_2}{2\varepsilon_\text{sol}+\varepsilon_2}
 \biggr) +  \frac{q_2^2 e^{2\Tilde a_2} \Tilde a_1^3}{(1+\Tilde a_2)^2} \biggl(\frac{1}{3} + \Bigl(1+\frac{1}{\Tilde R_{1 2}}\Bigr)^2 
 \frac{\varepsilon_\text{sol}-\varepsilon_1}{2\varepsilon_\text{sol}+\varepsilon_1}
 \biggr)\biggr] . \label{FLL_approx_E2}
\end{gather}
The right-hand side of \eqref{energy_N_monopoles_2_screened_legendre_1} boils down to the following many-body expression in the $\kappa\to0$ limit:
\begin{equation}
\label{E2_point_charges_N_spheres_s_kappa_zero}
\mathcal E^{(2)}\bigr|_{\kappa\to0} = \frac{-1}{8\pi\varepsilon_0\varepsilon_\text{sol}}\sum_{i=1}^N q_i \sum_{j=1,\, j\ne i}^N \, \sum_{p=1,\, p\ne j}^N q_p (\varepsilon_j-\varepsilon_\text{sol}) \sum_{l=1}^{+\infty}\frac{l a_j^{2 l+1}}{(l\varepsilon_j+(l+1)\varepsilon_\text{sol}) R_{i j}^{l+1} R_{j p}^{l+1}} P_l(\cos\gamma_{j i,j p}) .
\end{equation}

For the $\ell=2$-screened force we arrive at $\mathbf F_i^{(2)} = \Breve{\mathbf F}_i^{(2)} + \mathring{\mathbf F}_i^{(2)}$ in \eqref{Force_expansion_vec_xyz_components_ell} with leading terms (see \textcolor{red}{\cite{supplem_pre_force}} for derivation / further details)
\begin{subequations}
\label{forces_double_screening_all_}
\begin{align}
 \Breve{\mathbf F}_i^{(2)} & = \frac{d_{3,i} d_{4,i} \kappa^2}{4\pi\varepsilon_0\varepsilon_\text{sol}}\sum_{j=1, j\ne i}^N \; \sum_{p=1, p\ne i}^N \frac{q_j q_p e^{\Tilde a_j + \Tilde a_p} e^{-\Tilde R_{i j}} e^{-\Tilde R_{i p}}}{(1+\Tilde a_j) (1+\Tilde a_p) \Tilde R_{i j} \Tilde R_{i p}}\Bigl( 1 + \frac{1}{\Tilde R_{i p}} \Bigr) \pmb{\mathfrak F}_{i j i p} -\frac{q_i e^{\Tilde a_i} d_{1,i} \kappa^2}{8\pi\varepsilon_0\varepsilon_\text{sol}} \label{monopoles_double_standard} \\
&\quad \times\sum_{j=1, j\ne i}^N \; \sum_{p=1, p\ne j}^N \frac{q_p e^{\Tilde a_p} e^{-\Tilde R_{i j}}e^{-\Tilde R_{j p}}}{(1+\Tilde a_p)\Tilde R_{i j}\Tilde R_{j p}}\biggl[\Bigl( 1 +\frac{1}{\Tilde R_{i j}}\Bigr) d_{0,j} \Hat{\mathbf R}_{i j} + \bigl\{\Hat{\mathbf R}_{j p}+\pmb{\mathfrak F}_{i j j p}\bigr\}\Bigl( 1+\frac{1}{\Tilde R_{j p}}\Bigr) d_{2,j} \biggr]+ \cdots, \notag\\
 \mathring{\mathbf F}_i^{(2)} & = \frac{-\kappa^2\Tilde a_i^2}{4\pi\varepsilon_0} \sum_{j=1, j\ne i}^N \; \sum_{p=1, p\ne i}^N \frac{q_j q_p e^{\Tilde a_j + \Tilde a_p} e^{-\Tilde R_{i j}} e^{-\Tilde R_{i p}}}{(1+\Tilde a_j) (1+\Tilde a_p) \Tilde R_{i j} \Tilde R_{i p}} \biggl[ \frac{e^{2\Tilde a_i}\Tilde a_i d_{1,i}^\prime}{1+\Tilde a_i}\Bigl(1+\frac{1}{\Tilde R_{i j}}\Bigr) \Hat{\mathbf R}_{i j} + \varepsilon_\text{sol}d_{3,i}^\prime d_{4,i}^\prime\Bigl(1+\frac{1}{\Tilde R_{i p}}\Bigr) \pmb{\mathfrak F}_{i j i p} \biggr] \label{monopoles_double_osmotic} \\
&\quad - \frac{q_i e^{\Tilde a_i} d_{1,i}^\prime \kappa^2 \Tilde a_i^2}{8\pi\varepsilon_0(1+\Tilde a_i)}\sum_{j=1, j\ne i}^N \; \sum_{p=1, p\ne j}^N \frac{q_p e^{\Tilde a_p}}{1+\Tilde a_p} \frac{e^{-\Tilde R_{i j}}}{\Tilde R_{i j}}\frac{e^{-\Tilde R_{j p}}}{\Tilde R_{j p}}\biggl[\Bigl(1+\frac{1}{\Tilde R_{i j}}\Bigr) d_{0,j} \Hat{\mathbf R}_{i j} + \bigl\{ \Hat{\mathbf R}_{j p} + \pmb{\mathfrak F}_{i j j p}  \bigr\}\Bigl(1+\frac{1}{\Tilde R_{j p}}\Bigr) d_{2,j} \biggr] + \cdots  , \notag
\end{align}
\end{subequations}
where vector $\pmb{\mathfrak F}_{i j k p} \mathrel{:=} \bigl(1+3\Tilde R_{i j}^{-1}+3\Tilde R_{i j}^{-2}\bigr)\bigl(3(\Hat{\mathbf R}_{i j}\cdot\Hat{\mathbf R}_{k p})\Hat{\mathbf R}_{i j}-\Hat{\mathbf R}_{k p}\bigr)$, $\imath$ is a complex unit, $d_{1,i}^\prime \mathrel{:=} \bigl((\varepsilon_i+2\varepsilon_\text{sol})(1+\Tilde a_i)+\varepsilon_\text{sol}\Tilde a_i^2\bigr)^{-1}$, $d_{1,i} \mathrel{:=} (\varepsilon_i+2\varepsilon_\text{sol}) d_{1,i}^\prime$, $d_{3,i}^\prime \mathrel{:=} e^{2\Tilde a_i}\Tilde a_i^3 d_{1,i}^\prime$, $d_{4,i}^\prime \mathrel{:=} \bigl((2\varepsilon_i+3\varepsilon_\text{sol})(\Tilde a_i^2+3\Tilde a_i+3)+\varepsilon_\text{sol}\Tilde a_i^2(1+\Tilde a_i)\bigr)^{-1}$, $d_{3,i} \mathrel{:=} (\varepsilon_i-\varepsilon_\text{sol})d_{3,i}^\prime$, $d_{4,i} \mathrel{:=} (2\varepsilon_i+3\varepsilon_\text{sol}) d_{4,i}^\prime$. In the particular case of two ($N=2$) $z$-aligned spheres and $i=1$ and $j=2$, relation \eqref{monopoles_double_standard} boils down~to 
\begin{equation}
\label{force_two_spheres_2scr_nonosmotic}
(\Breve{\mathbf{F}}_i^{(2)})_z = \frac{q_i^2\kappa^2 e^{2\Tilde a_i - 2\Tilde R} d_{1,i}}{8\pi\varepsilon_0\varepsilon_\text{sol}(1+\Tilde a_i)\Tilde R^2}\!\Bigl(\!1\!+\!\frac{1}{\Tilde R}\!\Bigr)\!\biggl(\!\!\Bigl(3\!+\!\frac{6}{\Tilde R}\!+\!\frac{6}{\Tilde R^2}\!\Bigr)d_{2,j} - \!d_{0,j}\!\!\biggr) + \frac{q_j^2\kappa^2 e^{2\Tilde a_j-2\Tilde R} d_{3,i} d_{4,i} }{2\pi\varepsilon_0\varepsilon_\text{sol}(1\!+\!\Tilde a_j)^2\Tilde R^2}\!\Bigl(\!1\!+\!\frac{1}{\Tilde R}\Bigr)\!\Bigl(\!1\!+\!\frac{3}{\Tilde R}\!+\!\frac{3}{\Tilde R^2}\Bigr) + \cdots .
\end{equation}
If in \eqref{force_two_spheres_2scr_nonosmotic} one assumes $\varepsilon_i=\varepsilon_\text{sol}$ (sphere $i$ is non-polarizable), $q_j=0$ (sphere $j$ is polarizable with zero charge), $\Tilde a_i \ll \Tilde R \ll 1$ and $\Tilde a_j \ll \Tilde R \ll 1$ (weak screening and large inter-particle separation of small particles), and expands factors $d_{0,j}$ and $d_{2,j}$ w.r.t.~$\Tilde a_j$ (similarly to the derivation of the FLL energy relation \eqref{FLL_approx_E2} above), then \eqref{force_two_spheres_2scr_nonosmotic} yields relation derived in 2016, in \cite[Eq.~(34)]{Derb2}: 
\begin{equation}
\label{force_two_spheres_2scr_nonosmotic_Derb}
(\Breve{\mathbf{F}}_i^{(2)})_z \approx \frac{q_i^2 a_j^3 e^{-2\kappa R}(1+\kappa R)}{4\pi\varepsilon_0\varepsilon_\text{sol} R^5}\Bigl(\frac{\varepsilon_j-\varepsilon_\text{sol}}{\varepsilon_j+2\varepsilon_\text{sol}}(2+2\kappa R+\kappa^2 R^2)-\frac{1}{3}\kappa^2 R^2\Bigr) .
\end{equation}

\paragraph{\textbf{Janus particles.}} For Janus particles we have $\sigma_{n m,i}^\text{f} = \frac{\sqrt{4\pi}}{\sqrt{2 n+1}}(-1)^m Y_n^{-m}(\theta_i^r, \varphi_i^r)\sigma_{n 0,i}^\text{f, can}$, where angles $(\theta_i^r,\varphi_i^r)$ define the orientation of axis $\mathbf z_i^\prime$ (see panel \textbf{d} of Fig.~\ref{4panels}) and $\sigma_{0 0,i}^\text{f, can} = \frac{q_{i,1}+q_{i,2}}{\sqrt{4\pi} a_i^2}$, $\sigma_{n m,i}^\text{f, can} = \frac{\sqrt{\pi}}{\sqrt{2 n+1}} \bigl(\!\bigl(P_{n-1}(\cos\theta_{i,1})-P_{n+1}(\cos\theta_{i,1})\bigr)\varsigma_{i,1} + (-1)^n \bigl(P_{n-1}(\cos\theta_{i,2})-P_{n+1}(\cos\theta_{i,2})\bigr)\varsigma_{i,2}\!\bigr)\delta_{m,0}$ (as $n>0$), $\varsigma_{i,1 (2)} \mathrel{:=} \frac{q_{i,1 (2)}}{2\pi a_i^2 (1-\cos\theta_{i,1 (2)})}$, $P_n$ are the Legendre polynomials. Then,
\begin{align}
\mathcal E^{(0)} & = \frac{1}{8\pi\varepsilon_0\varepsilon_\text{sol}}  \sum_{i=1}^N \frac{q_i^2}{(1+\Tilde a_i) a_i} + \frac{3}{8\pi\varepsilon_0}\sum_{i=1}^N\frac{(1+\Tilde a_i)d_{1,i}^\prime}{a_i^3} \mathbf p_i \cdot \mathbf p_i + \; \cdots \, ;  \qquad \mathcal E^{(1)} = \frac{1}{8\pi\varepsilon_0\varepsilon_\text{sol}}\sum_{i=1}^N \sum_{j=1, j\ne i}^N \frac{q_i q_j e^{\Tilde a_i + \Tilde a_j} e^{-\Tilde R_{i j}}}{(1+\Tilde a_i) (1+\Tilde a_j) R_{i j}} \notag \\ 
&\qquad - \frac{3}{8\pi\varepsilon_0} \sum_{i=1}^N \sum_{j=1, j\ne i}^N \frac{q_i e^{\Tilde a_i + \Tilde a_j} d_{1,j}^\prime e^{-\Tilde R_{i j}}}{(1+\Tilde a_i) R_{i j}}\Bigl(\kappa+\frac{1}{R_{i j}}\Bigr) \mathbf p_j\cdot\Hat{\mathbf R}_{i j} + \frac{3}{8\pi\varepsilon_0} \sum_{i=1}^N \sum_{j=1, j\ne i}^N \frac{q_j e^{\Tilde a_i + \Tilde a_j} d_{1,i}^\prime e^{-\Tilde R_{i j}}}{(1+\Tilde a_j) R_{i j}}\Bigl(\kappa+\frac{1}{R_{i j}}\Bigr) \mathbf p_i\cdot\Hat{\mathbf R}_{i j} \label{Janus_expr} \\
&\qquad + \frac{9\varepsilon_\text{sol}}{8\pi\varepsilon_0} \sum_{i=1}^N \sum_{j=1, j\ne i}^N e^{\Tilde a_i + \Tilde a_j} d_{1,i}^\prime d_{1,j}^\prime \biggl[\!\Bigl(\frac{\kappa}{R_{i j}}+\frac{1}{R_{i j}^2}\Bigr) \mathbf p_i \cdot \mathbf p_j - \Bigl(\kappa^2+\frac{3\kappa}{R_{i j}}+\frac{3}{R^2_{i j}}\Bigr)(\mathbf p_i\cdot\Hat{\mathbf R}_{i j})(\mathbf p_j\cdot\Hat{\mathbf R}_{i j}) \biggr] \frac{e^{-\Tilde R_{i j}}}{R_{i j}} + \; \cdots \, . \notag
\end{align}

\end{widetext}

\bibliography{LinPaper}

@PREAMBLE{
 "\providecommand{\noopsort}[1]{}" 
 # "\providecommand{\singleletter}[1]{#1}%" 
}

@article{supplem_pre,
    author    = {S. V. Siryk and W. Rocchia},
    title     = {Many interacting particles in solution. {I}. {S}creening-ranged expansions of electrostatic potential and energy},
    journal   = {arXiv preprint},
    year      = {2025},
    volume    = {},
    number    = {},
    pages     = {},
    eprint    = {2512.08407},
    archivePrefix = {arXiv},
    doi       = {https://doi.org/10.48550/arXiv.2512.08407},
}

@article{supplem_pre_force,
    author    = {S. V. Siryk and W. Rocchia},
    title     = {Many interacting particles in solution. {II}. {S}creening-ranged expansion of electrostatic forces},
    journal   = {arXiv preprint},
    year      = {2025},
    volume    = {},
    number    = {},
    pages     = {},
    eprint    = {2512.08682},
    archivePrefix = {arXiv},
    doi       = {https://doi.org/10.48550/arXiv.2512.08682},
}

@article{supplem_pre_math,
    author    = {S. V. Siryk and W. Rocchia},
    title     = {Many interacting particles in solution. {III}. {S}pectral analysis of the associated {N}eumann--{P}oincar\'e-type operators},
    journal   = {arXiv preprint},
    year      = {2025},
    volume    = {},
    number    = {},
    pages     = {},
    eprint    = {2512.08684},
    archivePrefix = {arXiv},
    doi       = {https://doi.org/10.48550/arXiv.2512.08684},
}

@misc{our_github_rep,
howpublished = {\url{https://github.com/concept-lab/Analytical_Electrostatics}},
title = {{GitHub} repository},
year = {2025},
}

@article{our_jcp,
author = {S. Siryk and A. Bendandi and A. Diaspro and W. Rocchia},
journal = {J. Chem. Phys.},
title = {Charged dielectric spheres interacting in electrolytic solution: a linearized {P}oisson-{B}oltzmann equation model},
year = {2021},
volume = {155},
pages = {114114},
issue = {11},
}

@article{our_jpcb,
author = {S. Siryk and W. Rocchia},
journal = {J. Phys. Chem. B},
title = {Arbitrary-Shape Dielectric Particles Interacting in the Linearized {P}oisson-{B}oltzmann Framework: An Analytical Treatment},
year = {2022},
volume = {126},
pages = {10400-10426},
issue = {49},
}

@article{SheinermanNorelHonig,
author = {F. B. Sheinerman and R. Norel and B. Honig},
journal = {Curr. Opin. Struct. Biol.},
pages = {153159},
title = {Electrostatic aspects of protein protein interactions},
volume = {10},
year = {2000},
}

@article{Yu3,
author = {Y.-K. Yu},
journal = {Phys. Rev. E},
pages = {052404},
title = {Electrostatics of charged dielectric spheres with application to biological systems. {III}. {R}igorous ionic screening at the {D}ebye-{H}\"uckel level},
volume = {102},
year = {2020},
issue = {5},
}

@article{BMP22,
author = {C. Chen and B. Yu and R. Yousefi and J. Iwahara and B.M. Pettitt},
journal = {J. Phys. Chem. B},
pages = {4543-4554},
title = {Assessment of the components of the electrostatic potential of proteins in solution: comparing experiment and theory},
volume = {126},
year = {2022},
}

@article{Yu2021,
author = {O. I. Obolensky and T. P. Doerr and Yi-Kuo Yu},
journal = {Eur. Phys. J. E},
pages = {129},
title = {Rigorous treatment of pairwise and many-body electrostatic interactions among dielectric spheres at the {D}ebye-{H}\"uckel level},
volume = {44},
year = {2021},
}

@article{Tabrizi1,
author = {A. M. Tabrizi and S. Goossens and A. M. Rahimi and C. D. Cooper and M. G. Knepley and J. P. Bardhan},
journal = {J. Chem. Theory Comput.},
number = {6},
pages = {2897-2914},
title = {Extending the solvation-layer interface condition continum electrostatic model to a linearized {P}oisson-{B}oltzmann solvent},
volume = {13},
year = {2017},
}

@article{RAH1,
author = {W. Rocchia and E. Alexov and B. Honig},
journal = {Journal of Physical Chemistry B},
number = {28},
pages = {6507-6514},
title = {Extending the applicability of the nonlinear {P}oisson-{B}oltzmann equation: multiple dielectric constants and multivalent ions},
volume = {105},
year = {2001},
}

@article{Silva1,
author = {G. M. Silva and X. Liang and G. M. Kontogeorgis},
journal = {J. Phys. Chem. B},
volume = {126},
pages = {4112-4131},
title = {Investigation of the limits of the linearized {P}oisson-{B}oltzmann equation},
year = {2022},
}

@incollection{MAR1,
  author      = {T. Markovich and D. Andelman and R. Podgornik},
  title       = {Charged membranes: {P}oisson-{B}oltzmann theory, the {DLVO} paradigm, and beyond},
  editor      = {C. R. Safinya and J. O. R\"adler},
  booktitle   = {Handbook of lipid membranes},
  publisher   = {CRC Press},
  address     = {Boca Raton},
  year        = {2021},
  pages       = {99-128},
  chapter     = {6},
}

@article{Kirkwood1934,
author = {J. G. Kirkwood},
journal = {Journal of Chemical Physics},
pages = {351-361},
title = {Theory of Solutions of Molecules Containing Widely Separated Charges with Special Application to Zwitterions},
volume = {2},
year = {1934},
}

@article{Alex1,
author = {S. Alexander and P. M. Chaikin and P. Grant and G. J. Morales and P. Pincus and D. J. Hone},
journal = {Journal of Chemical Physics},
pages = {5776-5781},
title = {Charge renormalization, osmotic pressure, and bulk modulus of colloidal crystals: Theory},
volume = {80},
year = {1984},
}

@article{Triz1,
author = {E. Trizac and L. Bocquet and M. Aubouy and H. H. von Grunberg},
journal = {Langmuir},
pages = {4027-4033},
title = {Alexander's prescription for colloidal charge renormalization},
volume = {19},
year = {2003},
issue = {9},
}

@article{Boon2015,
author = {N. Boon and G. I. Guerrero-Garcia and R. van Roij and M. O. de la Cruz},
journal = {Proc. Natl. Acad. Sci. U.S.A.},
pages = {9242-9246},
title = {Effective charges and virial pressure of concentrated macroion solutions},
volume = {112},
year = {2015},
number = {30},
}

@article{ST1,
author = {L. Samaj and E. Trizac},
journal = {J. Phys. A},
pages = {265003},
title = {Effective charge of cylindrical and spherical colloids immersed in an electrolyte: the quasi-planar limit},
volume = {48},
year = {2015},
issue={26},
}

@article{TrizPRL,
author = {E. Trizac and L. Bocquet and M. Aubouy},
journal = {Phys. Rev. Lett.},
pages = {248301},
title = {Simple Approach for Charge Renormalization in Highly Charged Macroions},
volume = {89},
year = {2002},
}

@article{SCW,
author = {S. D. Search and C. D. Cooper and E. van{'}t Wout},
journal = {J. Comput. Chem.},
pages = {674-691},
title = {Towards optimal boundary integral formulations of the {P}oisson-{B}oltzmann equation for molecular electrostatics},
volume = {43},
year = {2022},
issue={10},
}

@article{WK2,
author = {L. Wilson and W. Geng and R. Krasny},
journal = {J. Phys. Chem. B},
pages = {7104-7113},
title = {{TABI-PB} 2.0: An improved version of the treecode-accelerated boundary integral {P}oisson-{B}oltzmann solver},
volume = {126},
year = {2022},
issue={37},
}

@article{JNQS,
author = {S. Nakov and E. Sobakinskaya and T. Renger and J. Kraus},
journal = {J. Comput. Chem.},
number = {26},
pages = {1832-1860},
title = {{ARGOS}: {A}n adaptive refinement goal-oriented solver for the linearized {P}oisson-{B}oltzmann equation},
volume = {42},
year = {2021},
}

@article{Derb2,
author = {I. N. Derbenev and A. V. Filippov and A. J. Stace and E. Besley},
journal = {J. Chem. Phys.},
issue = {8},
pages = {084103},
title = {Electrostatic interactions between charged dielectric particles in an electrolyte solution},
volume = {145},
year = {2016},
}

@article{Derb4,
author = {I. N. Derbenev and A. V. Filippov and A. J. Stace and E. Besley},
journal = {Soft Matter},
pages = {5480-5487},
title = {Electrostatic interactions between charged dielectric particles in an electrolyte solution: constant potential boundary conditions},
volume = {14},
year = {2018},
}

@article{Fish,
author = {M. E. Fisher and Y. Levin and X. Li},
journal = {J. Chem. Phys.},
pages = {2273-2282},
title = {The interaction of ions in an ionic medium},
volume = {101},
year = {1994},
issue = {3},
}

@article{FinBar2016,
author = {S. Finlayson and P. Bartlett},
journal = {J. Chem. Phys.},
pages = {034905},
title = {Non-additivity of pair interactions in charged colloids},
volume = {145},
year = {2016},
issue = {3},
}

@article{LLF,
author = {X. Li and Y. Levin and M. Fisher},
journal = {Europhys. Lett.},
pages = {683-688},
title = {Cavity forces and criticality in electrolytes},
volume = {26},
year = {1994},
number = {9},
}

@article{FBYJBH,
  title = {{PB-AM}: An Open-Source, Fully Analytical Linear {P}oisson-{B}oltzmann Solver},
  author = {L. E. Felberg and D. H. Brookes and E.-H. Yap and E. Jurrus and N. A. Baker and T. Head-Gordon},
  journal = {J. Comput. Chem.},
  volume = {38},
  pages = {1275-1282},
  year = {2017},
  issue = {15},
}

@article{Yu2019,
author = {Y.-K. Yu},
journal = {Phys. Rev. E},
pages = {012401},
title = {Electrostatics of charged dielectric spheres with application to biological systems. {II}. {A} formalism bypassing {W}igner rotation matrices},
volume = {100},
issue = {1},
year = {2019},
}

@article{McClurg,
author = {R. B. McClurg and C. F. Zukoski},
journal = {J. Colloid Interface Sci.},
issue = {2},
pages = {529-542},
title = {The electrostatic interaction of rigid, globular proteins with arbitrary charge distributions},
volume = {208},
year = {1998},
}

@article{Chan,
author = {H.-K. Chan},
journal = {Journal of Electrostatics},
pages = {103435},
title = {A theory for like-charge attraction of polarizable ions},
volume = {105},
year = {2020},
}

@article{Qin2019,
author = {J. Qin},
journal = {Soft Matter},
pages = {2125-2134},
title = {Charge polarization near dielectric interfaces and the multiple-scattering formalism},
volume = {15},
year = {2019},
}

@article{GJLX,
author = {Z. Gan and S. Jiang and E. Luijten and Z. Xu},
journal = {SIAM J. Sci. Comput.},
number = {3},
pages = {B375-B395},
title = {A hybrid method for systems of closely spaced dielectric spheres and ions},
volume = {38},
year = {2016},
}

@article{LinQS,
author = {E. Lindgren and C. Quan and B. Stamm},
journal = {J. Chem. Phys.},
pages = {044901},
title = {Theoretical analysis of screened many-body electrostatic interactions between charged polarizable particles},
volume = {150},
year = {2019},
}

@article{Ether2018,
author = {D. S. Ether and F. S. S. Rosa and D. M. Tibaduiza and L. B. Pires and R. S. Decca and P. A. Maia Neto},
journal = {Phys. Rev. E},
pages= {022611},
number = {2},
title = {Double-layer force suppression between charged microspheres},
volume = {97},
year = {2018},
}

@article{LCSB,
author = {E. B. Lindgren and H.-K. Chan and A. J. Stace and E. Besley},
journal = {Phys. Chem. Chem. Phys.},
pages = {5883-5895},
title = {Progress in the theory of electrostatic interactions between charged particles},
volume = {18},
year = {2016},
}

@article{FolOnu,
author = {D. E. Folescu and A. V. Onufriev},
journal = {ACS Omega},
pages = {26123-26136},
title = {A closed-form, analytical approximation for apparent surface charge and electric field of molecules},
volume = {7},
year = {2022},
issue = {30},
}

@article{DerbFil2017,
author = {A. V. Filippov and I. N. Derbenev and A. A. Pautov and M. M. Rodin},
journal = {J. Exp. Theor. Phys.},
number = {3},
pages = {518-529},
title = {Electrostatic interaction of macroparticles in a plasma in the strong screening regime},
volume = {125},
year = {2017},
}

@article{FilStar_jpcb,
author = {A. Filippov and V. Starov},
journal = {J. Phys. Chem. B},
number = {29},
pages = {6562-6572},
title = {Interaction of Nanoparticles in Electrolyte Solutions},
volume = {127},
year = {2023},
}

@article{BesleyACR,
author = {E. Besley},
journal = {Acc. Chem. Res.},
number = {17},
pages = {2267-2277},
title = {Recent Developments in the Methods and Applications of Electrostatic Theory},
volume = {56},
year = {2023},
}

@article{DY_2006,
author = {T.P. Doerr and Y.-K. Yu},
journal = {Phys. Rev. E},
pages = {061902},
title = {Electrostatics of charged dielectric spheres with application to biological systems},
volume = {73},
year = {2006},
issue = {6},
}

@article{Hassan_Stamm_jctc,
author = {M. Hassan and C. Williamson and J. Baptiste and S. Braun and A.J. Stace and E. Besley and B. Stamm},
journal = {J. Chem. Theory Comput.},
pages = {6281-6296},
title = {Manipulating Interactions between Dielectric Particles with Electric Fields: A General Electrostatic Many-Body Framework},
volume = {18},
issue = {10},
year = {2022},
}

@article{SchlaichHolm,
author = {A. Schlaich and S. Tyagi and S. Kesselheim and M. Sega and C. Holm},
journal = {Eur. Phys. J. E},
pages = {80},
title = {Renormalized charge and dielectric effects in colloidal interactions: a numerical solution of the nonlinear {P}oisson-{B}oltzmann equation for unknown boundary conditions},
volume = {46},
year = {2023},
}

@article{BritoDenton,
author = {M. E. Brito and G. N\"agele and A. R. Denton},
journal = {J. Chem. Phys.},
pages = {204904},
title = {Effective interactions, structure, and pressure in charge-stabilized colloidal suspensions: Critical assessment of charge renormalization methods},
volume = {159},
year = {2023},
issue = {20},
}

@article{Jha1,
author = {A. Jha and M. Nottoli and A. Mikhalev and C. Quan and B. Stamm},
journal = {J. Chem. Phys.},
pages = {104105},
title = {Linear scaling computation of forces for the domain-decomposition linear {P}oisson-{B}oltzmann method},
volume = {158},
year = {2023},
issue = {10},
}

@article{Lindgren_jcis,
author = {E. B. Lindgren and H. Avis and A. Miller and B. Stamm and E. Besley and A. J. Stace},
journal = {J. Colloid Interface Sci.},
pages = {458-466},
title = {The significance of multipole interactions for the stability of regular structures composed from charged particles},
volume = {663},
year = {2024},
}

@article{AddisonSmithCooper2023,
author = {I. Addison-Smith and H. V. Guzman and C. D. Cooper},
journal = {Journal of Chemical Theory and Computation},
issue = {10},
pages = {2996-3006},
title = {Accurate Boundary Integral Formulations for the Calculation of Electrostatic Forces with an Implicit-Solvent Model},
volume = {19},
year = {2023},
}

@article{BoWangjcp,
author = {B. Wang and W. Zhang and W. Cai},
journal = {J. Comp. Phys.},
pages = {110379},
title = {Fast multipole method for 3-{D} {P}oisson-{B}oltzmann equation in layered electrolyte-dielectric media},
volume = {439},
year = {2021},
}

@book{Blossey2023,
author = {R. Blossey},
title = {The {P}oisson-{B}oltzmann Equation},
address = {Cham, Switzerland},
year = {2023},
publisher = {Springer Nature}
}

@article{KMRR,
author = {P. Khunpetch and A. Majee and H. Ruixuan and R. Podgornik},
journal = {Phys. Rev. E},
pages = {024402},
title = {Curvature effects in interfacial acidity of amphiphilic vesicles},
volume = {108},
year = {2023},
}

@article{RMDP,
author = {H. Ruixuan and A. Majee and J. Dobnikar and R. Podgornik},
journal = {The European Physical Journal E},
pages = {115},
title = {Electrostatic interactions between charge regulated spherical macroions},
volume = {46},
year = {2023},
}

@article{Krishnan2017,
author = {M. Krishnan},
journal = {J. Chem. Phys.},
pages = {205101},
title = {A simple model for electrical charge in globular macromolecules and linear polyelectrolytes in solution},
volume = {146},
year = {2017},
issue = {20},
}

@article{BSBBC,
author = {M. Bosy and M. W. Scroggs and T. Betcke and E. Burman and C. D. Cooper},
journal = {J. Comput. Chem.},
pages = {787-797},
title = {Coupling finite and boundary element methods to solve the {P}oisson-{B}oltzmann equation for electrostatics in molecular solvation},
volume = {45},
year = {2024},
issue = {11},
}

@article{LSB2018,
author = {E. B. Lindgren and A. J. Stace and E. Polack and Y. Maday and B. Stamm and E. Besley},
journal = {Journal of Computational Physics},
pages = {712-731},
title = {An integral equation approach to calculate electrostatic interactions in many-body dielectric systems},
volume = {371},
year = {2018},
}

@article{HasStammI,
author = {M. Hassan and B. Stamm},
journal = {ESAIM: M2AN},
pages = {S65-S102},
title = {An integral equation formulation of the {N}-body dielectric spheres problem. part {I}: numerical analysis},
volume = {55},
year = {2021},
}

@article{GXFQ,
author = {K. S. Gustafson and G. Xu and K. F. Freed and J. Qin},
journal = {J. Chem. Phys.},
pages = {064908},
title = {Image method for electrostatic energy of polarizable dipolar spheres},
volume = {147},
year = {2017},
}

@article{QPF,
author = {J. Qin and J. de Pablo and K. F. Freed},
journal = {J. Chem. Phys.},
pages = {124903},
title = {Image method for induced surface charge from many-body system of dielectric spheres},
volume = {145},
year = {2016},
}

@article{BarrLui_2014,
author = {K. Barros and E. Luijten},
journal = {Phys. Rev. Lett.},
pages = {017801},
title = {Dielectric Effects in the Self-Assembly of Binary Colloidal Aggregates},
volume = {113},
year = {2014},
}

@article{CurLui_2021,
author = {Tine Curk and Erik Luijten},
journal = {Phys. Rev. Lett.},
pages = {138003},
title = {Charge Regulation Effects in Nanoparticle Self-Assembly},
volume = {126},
year = {2021},
}

@article{CMWA,
author = {K. Coshic and C. Maffeo and D. Winogradoff and A. Aksimentiev},
journal = {Nature},
pages = {905-914},
title = {The structure and physical properties of a packaged bacteriophage particle},
volume = {627},
year = {2024},
}

@article{GraafBoon2012,
author = {J. de Graaf and N. Boon and M. Dijkstra and R. van Roij},
journal = {J. Chem. Phys.},
pages = {104910},
title = {Electrostatic interactions between {J}anus particles},
volume = {137},
year = {2012},
issue = {10},
}

@article{BoonJanus2010,
author = {N. Boon and E. Carvajal Gallardo and S. Zheng and E. Eggen and M. Dijkstra and R. van Roij},
journal = {J. Phys.: Condens. Matter},
pages = {104104},
title = {Screening of heterogeneous surfaces: charge renormalization of {J}anus particles},
volume = {22},
year = {2010},
}

@article{BSL2014,
author = {K. Barros and D. Sinkovits and E. Luijten},
journal = {J. Chem. Phys.},
pages = {064903},
title = {Efficient and accurate simulation of dynamic dielectric objects},
volume = {140},
year = {2014},
}

@article{Freed2014,
author = {K. F. Freed},
journal = {J. Chem. Phys.},
pages = {034115},
title = {Perturbative many-body expansion for electrostatic energy and field for system of polarizable charged spherical ions in a dielectric medium},
volume = {141},
year = {2014},
issue = {3},
}

@article{GWJXL,
author = {Z. Gan and Z. Wang and S. Jiang and Z. Xu and E. Luijten},
journal = {J. Chem. Phys.},
issue = {2},
pages = {024112},
title = {Efficient dynamic simulations of charged dielectric colloids through a novel hybrid method},
volume = {151},
year = {2019},
}

@article{JPCBSIYR,
author = {W. Yang and W. Rocchia},
title = {Biomolecular Electrostatic Phenomena: An Evergreen Field},
journal = {J. Phys. Chem. B},
volume = {127},
number = {18},
pages = {3979-3981},
year = {2023},
}

@article{DiFlorio2025NextGenPB,
    author    = {Di Florio, Vincenzo and Ansalone, Patrizio and Siryk, Sergii V. and Decherchi, Sergio and de Falco, Carlo and Rocchia, Walter},
    title     = {{NextGenPB}: An analytically-enabled super resolution tool for solving the {P}oisson-{B}oltzmann equation featuring local (de)refinement},
    journal   = {Computer Physics Communications},
    year      = {2025},
    volume    = {317},
    pages     = {109816},
}

@article{DGC_softmat,
author = {Y. Duan and Z. Gan and  H.-K. Chan},
journal = {Soft Matter},
pages = {1860-1872},
title = {Mechanisms of electrostatic interactions between two charged dielectric spheres inside a polarizable medium: an effective-dipole analysis},
volume = {21},
year = {2025},
}

@article{RPSA_2024,
author = {J. M. Randazzo and R. Della Picca and A. J. Sarsa and L. U. Ancarani},
journal = {Proc. R. Soc. A},
pages = {20240357},
title = {Spherical image potentials for an {N}-charged particle system},
volume = {480},
year = {2024},
issue = {2300},
}

@article{DuanGan2025,
author = {Y. Duan and Z. Gan},
title = {Quantitative Theory for Critical Conditions of Like-Charge Attraction Between Polarizable Spheres},
journal = {J. Chem. Theory Comput.},
year = {2025},
volume = {21},
issue = {6},
pages = {2822-2828},
}

@article{GLCBB_2025,
author = {A. Gnidovec and E. Locatelli and S. Copar and A. Bozic and E. Bianchi},
journal = {Nature Communications},
pages = {4277},
title = {Anisotropic {DLVO}-like interaction for charge patchiness in colloids and proteins},
volume = {16},
year = {2025},
}

@article{Ponce2024,
author = {Gustavo Ch. {Ponce de Leon} and G. Palasantzas},
journal = {Phys. Rev. Lett.},
pages = {246202},
title = {Effects of Patch Potentials in Electrostatic Double-Layer Forces},
volume = {133},
issue = {24},
year = {2024},
}

@article{FEM-BEM_diffuse_interface,
  title={Some Challenges of Diffused Interfaces in Implicit-Solvent Models},
  author={Guerrero-Montero, Mauricio and Bosy, Micha{\l} and Cooper, Christopher D},
  journal={Journal of Computational Chemistry},
  volume={46},
  number={3},
  pages={e70036},
  year={2025},
}

@article{CheDzu2008,
author = {J. Che and J. Dzubiella and B. Li and J. A. McCammon},
journal = {J. Phys. Chem. B},
pages = {3058-3069},
title = {Electrostatic free energy and its variations in implicit solvent models},
volume = {112},
issue = {10},
year = {2008},
}

@article{MajOsh2025,
author = {P. S. Majee and H. Ohshima},
journal = {Phys. Rev. E},
pages = {055407},
title = {Analytical study on diffusiophoresis of soft particles: Role of surface-charge-mobility-dependent hydrophobic core},
volume = {111},
issue = {5},
year = {2025},
}

@article{DGKF_2025,
author = {Marc Domingo and Horacio V. Guzman and Matej Kanduc and Jordi Faraudo},
journal = {J. Chem. Inf. Model.},
pages = {240-251},
title = {Electrostatic Interaction between {SARS-CoV-2} and Charged Surfaces: Spike Protein Evolution Changed the Game},
volume = {65},
issue = {1},
year = {2025},
}

@article{Cisneros_ChemRev,
author = {G. A. Cisneros and M. Karttunen and P. Ren and C. Sagui},
journal = {Chem. Rev.},
pages = {779-814},
title = {Classical Electrostatics for Biomolecular Simulations},
volume = {114},
year = {2014},
issue = {1},
}

@article{JanNetz09,
author = {J. Janecek and R. R. Netz},
journal = {J. Chem. Phys.},
volume = {130},
pages = {074502},
title = {Effective screening length and quasiuniversality for the restricted primitive model of an electrolyte solution},
year = {2009},
issue = {7},
}
\end{document}